\documentstyle[prl,aps,epsf,multicol,tabularx]{revtex}
\tighten
\begin{document}
\draft
\title{Positive Current Correlations in Gated Mesoscopic Conductors.}
\title{Charge fluctuations generate positive current-current fluctuations}
\title{Coulomb induced positive current-current correlations in normal
conductors}
\author{Andrew M. Martin and Markus B\"uttiker}
\address{D\'epartement de Physique Th\'eorique, Universit\'e de Gen\`eve,
CH-1211 Gen\`eve 4, Switzerland}
\date{\today}
\maketitle
\begin{abstract}
In the white-noise limit current correlations measured at
different contacts of a mesoscopic conductor are negative due to
the antisymmetry of the wave function (Pauli principle). We show
that current fluctuations at capacitive contacts induced via the
long range Coulomb interaction due to charge fluctuations in the
mesoscopic sample can be {\it positively} correlated. The positive
correlations are a consequence of the extension of the
wave-functions into areas near both contacts. As an example we
investigate in detail a quantum point contact in a high magnetic
field under conditions in which transport is along an edge state.
\end{abstract}
\pacs{Pacs numbers: 72.10.-d, 72.70+m}
\begin{multicols}{2}
\narrowtext
Fluctuations in beams of photons have long been of interest in
optics. In 1954 Hanbury Brown and Twiss \cite{Experiment0} showed
that an investigation of the correlations of light of a star
permits the determination of the diameter of the star. Electrical
analogs to this effect have long been suggested but due to the
interaction between carriers, it is difficult to achieve the
necessary degeneracy in a vacuum beam experiment. In electrical
conductors however, the situation is fundamentally different,
since at low temperatures, high filling factors are easy to
achieve. Current-current correlations in multi-terminal conductors
were analyzed theoretically using both quantum theories and
classical approaches \cite{Buttiker1,tmrl,corr,blanter}. It can
easily be demonstrated that in the white-noise limit,
current-current correlations in phase-coherent conductors are
negative \cite{Buttiker1,texier}. Indeed, recent experiments by
Henny et al. \cite{Experiment1}, Oliver et al. \cite{Experiment2}
and Oberholzer et al. \cite{Experiment3} have demonstrated
negative correlations in good agreement with theory. The fermionic
character of current fluctuations demonstrated in these
experiments reflects the independent quasi-particle transport in
the low frequency white noise limit \cite{note1}. However, such a
simple situation cannot be expected to hold over a wide range of
frequencies. It is well known, that time-dependent transport in
conductors is collective due the long range Coulomb interaction,
which leads to bosonic excitations (plasmons) and that
consequently we expect a different (bosonic) behavior of
current-current fluctuations. In this work, we demonstrate, that
already at (in principle) arbitrarily low frequencies, it is
possible to find current-correlations which are {\it positively}
correlated in purely normal conductors. To achieve this, we
investigate current fluctuations and their correlations at
contacts which are purely capacitively coupled to a conductor.
Current at these contacts is a consequence of charge fluctuations
in the mesoscopic conductor and not as in the theories and
experiments discussed above a consequence of carriers transmitted
or reflected into different contacts.

We initially consider the general system shown in Fig. 1, where we
have a mesoscopic conductor in contact with two ideal leads ($1$
and $2$) and Coulomb coupled to two macroscopic gates ($A$ and
$B$). We consider the scenario where the two gates are Coulomb
coupled to separate regions of the conductor ($\Omega_A$ and
$\Omega_B$), and assume that there is no direct Coulomb coupling
between regions $\Omega_A$ and $\Omega_B$.  It will be apparent
how such effects could be taken into account. Having shown how it
is possible to calculate the charge correlations between the two
gates ($A$ and $B$) we proceed to consider four specific examples,
for a Quantum Point Contact (QPC) in a high magnetic field,
demonstrating that for these systems the current correlations can
be both positive and negative.

Within the context of scattering theory the charging and charge
fluctuation phenomena in gated mesoscopic conductors have already
been investigated \cite{Buttiker2,Pedersen}. The work in this
Letter develops this understanding for structures with more than
one macroscopic gate coupled to the mesoscopic conductor. It has
already  been shown \cite{Pedersen} that for a mesoscopic
structure Coulomb coupled to a single macroscopic gate the
fluctuations in the charge in the mesoscopic region induce charge
fluctuations in a macroscopic gate which is placed in close
proximity to the mesoscopic region. The dynamics of the system are
governed by two quantities, the charge relaxation resis-
%
\begin{figure}
\narrowtext
\epsfxsize=7cm \centerline{\epsffile{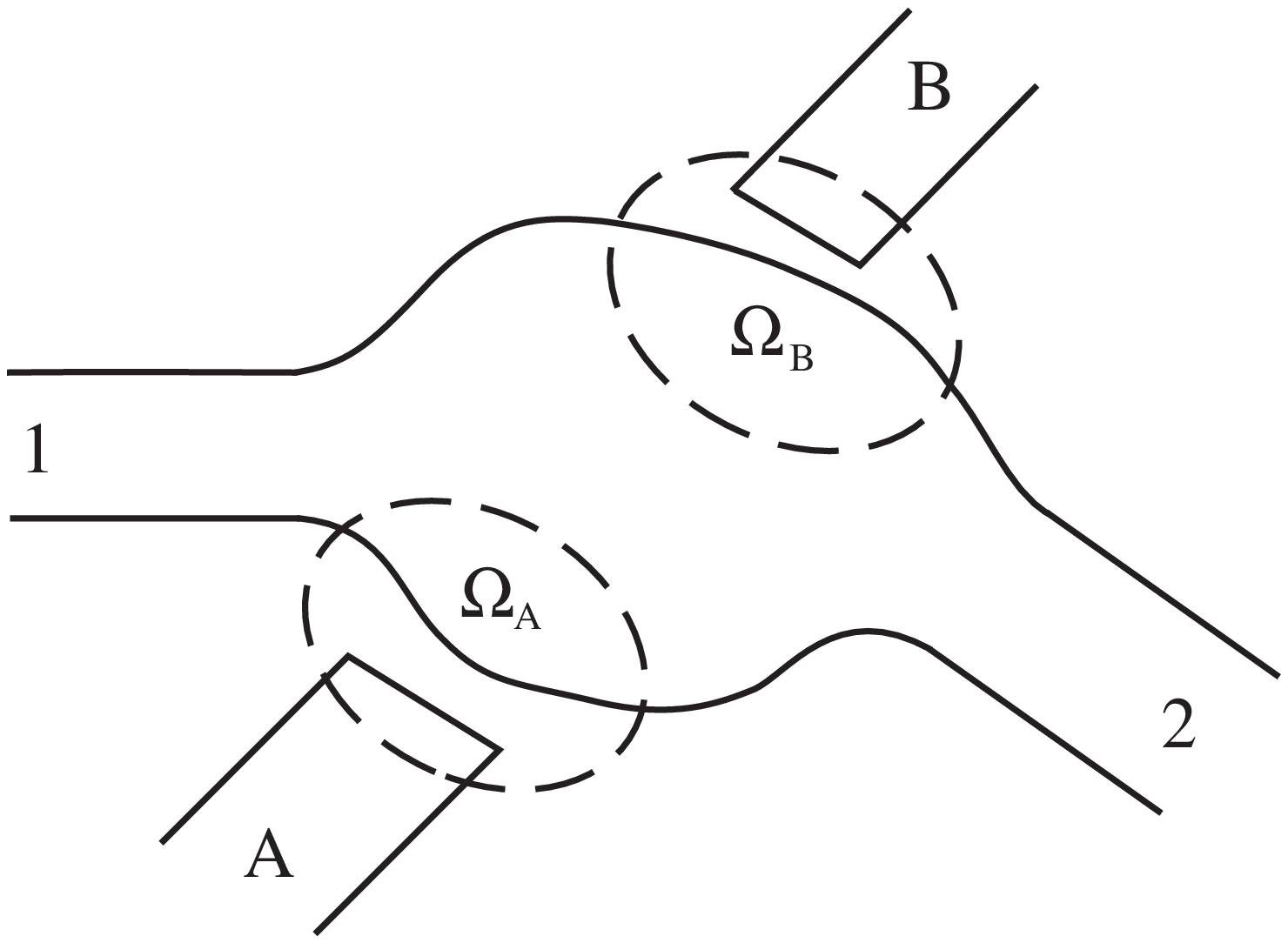}} \vspace{0.3cm}
\caption{ \label{general} Conductor connected via ideal leads ($1$
and $2$) to reservoirs with spatially separate regions
capacitively coupled to gates ($A$ and $B$).}
\end{figure}
\noindent  tance of the structure $R$ and the electrochemical
capacitance of the structure $C_{\mu}$. $RC_{\mu}$, in analogy
with macroscopic systems, denotes the charge relaxation time of
the conductor. For a mesoscopic system, at equilibrium, Coulomb
coupled to a single macroscopic gate, in the zero temperature
limit, the current fluctuation spectrum in the gate, to leading
order in the frequency, is given by \cite{Pedersen}
$S_{II}^q(\omega)=2 \omega^2 \hbar |\omega|C_{\mu}^2 R_q$, where
$R_q$ is the equilibrium charge relaxation resistance. Out of
equilibrium and at zero temperature it has been shown to leading
order in the applied voltage that \cite{Pedersen}
$S_{II}^V(\omega)=2 \omega^2 |eV|C_{\mu}^2 R_V$, where we call
$R_V$ the Schottky resistance to emphasize that the charge
fluctuations are associated with shot noise. In the presence of
more than one gate we show below that the equivalent spectrum is
now given by
\begin{equation}
S_{I_{\alpha}I_{\beta}}^q(\omega)= 2 \omega^2 \hbar
|\omega|C_{\mu}^{\alpha}C_{\mu}^{\beta} R_q^{\alpha \beta},
\label{SII1}
\end{equation}
\begin{equation}
S_{I_{\alpha}I_{\beta}}^V(\omega)= 2 \omega^2
|eV|C_{\mu}^{\alpha}C_{\mu}^{\beta} R_V^{\alpha \beta},
\label{SII2}
\end{equation}
where the sub and superscripts $\alpha$ and $\beta$ specify the
gates in the system. For example if $\alpha=\beta$ then Eq.
(\ref{SII1}) defines the equilibrium current fluctuation spectrum
in gate $\alpha$, whereas if $\alpha \ne \beta$ Eq. (\ref{SII1})
defines the equilibrium current correlations between gates
$\alpha$ and $\beta$.

Consider a mesoscopic conductor attached to $L$ normal ideal
leads. This conductor can be described by a scattering matrix,
with elements $s_{\gamma \delta}$, which relates electron
amplitudes incident from contact $\delta$ to outgoing amplitudes
in contact $\gamma$. If we let regions of the mesoscopic conductor
be capacitively coupled to gates, as depicted in Fig. 1, the
current fluctuations induced in a given gate are related to the
charge fluctuations in the region of the conductor which is
Coulomb coupled to the gate. If we have, as in Fig. 1, more than
one gate then we need to calculate the charge fluctuations in
separate regions of the conductor, allowing us to then calculate
both the charge fluctuations and the correlations of the currents
in the gates. The bare charge fluctuations in a region
$\Omega_{\alpha}$ are governed by the density of states matrix
\cite{Buttiker2}, whose elements are
\begin{equation}
{\cal N}^{(\eta)}_{\delta \gamma}({\bbox r}) = \frac{-1}{2 \pi
i}\sum_{\nu} \left[ s^{\star}_{\nu \delta}(E, U({\bbox r}))
\frac{\delta s_{\nu \gamma}(E, U({\bbox r}))}{e\delta U({\bbox
r})} \right] \label{d0}
\end{equation}
where ${\bbox r}$ is in the volume $\Omega_{\eta}$ and $U({\bbox
r})$ is the electrostatic potential at position ${\bbox r}$. For
example ${\cal N}^{(A)}_{1 2}({\bbox r})$ is the electron density,
at position $\bf{r}$ in volume $\Omega_A$, associated with {\it
two electron current amplitudes} incident from contacts $1$ and
$2$. ${\cal N}^{(\eta)}_{\gamma \gamma}({\bbox r})$ are elements
of a generalized {\it local} Wigner-Smith time-delay matrix
\cite{ws}. The explicit relation of the charge operator to local
wave functions is given in \cite{Buttiker2} and a detailed
derivation is found in Ref. \cite{mb12}.

The fluctuations of the bare charge in a region $\Omega_{\eta}$
can be found from the charge operator \cite{Pedersen} $e
\hat{{\cal N}}_{(\eta)}$ given by
\begin{eqnarray}
\hat{{\cal N}}_{\eta}(\omega) &=& \sum_{\delta \gamma}
\int_{\Omega_{\eta}} d^3 {\bbox r} \int dE \, \,
\hat{a}^{\dagger}_{\delta}(E) \nonumber \\
 &\times& {\cal N}_{\delta \gamma}^{(\eta)}
({\bbox r}; E, E+\hbar \omega) \hat{a}_{\gamma}(E+\hbar \omega),
\label{Noper}
\end{eqnarray}
where the zero-frequency limit of ${\cal N}_{\delta
\gamma}^{(\eta)} ({\bbox r}; E, E+\hbar \omega)$ is given by Eq.
(\ref{d0}). In the above equation $\hat{a}^{\dagger}_{\delta}(E)$
creates an incoming electron in lead $\delta$. The true charge
fluctuations must be obtained by taking into account the Coulomb
interaction.

Given the above equations we now want to consider the system shown
in Fig. 1 and to calculate charge fluctuations in the two regions
$\Omega_A$ and $\Omega_B$, hence enabling us to deduce the charge
in each of the two gates. The first step is to find an expression
for the charge operator in the two regions. Assuming that the
geometrical capacitances which couple the two gates to two
separate regions of the mesoscopic conductor dominate all other
capacitances we can express the charge in the two regions of the
conductor in two ways. First with the help of the potential
operators for the two regions, $\hat{U}_{A(B)}$, we have
\begin{equation}
\hat{Q}_{A}= C_{A} \hat{U}_{A}+C(\hat{U}_A-\hat{U}_B),
\end{equation}
\begin{equation}
\hat{Q}_{B}= C_{B} \hat{U}_{B}+C(\hat{U}_B-\hat{U}_A),
\end{equation}
where $C$ is the capacitive coupling between regions $\Omega_A$
and $\Omega_B$. The two above equations assume that the gates, $A$
and $B$ are macroscopic and have no dynamics of their own. We can
also determine the charge in a region $\hat{Q}_{A(B)}$ as the sum
of the bare charge fluctuations in the region of interest $e
\hat{{\cal N}}_{A(B)}$ and the induced charges generated by a
fluctuating induced electrical potential in the same region.
Within the random phase approximation the induced charges are
proportional to the average frequency dependent density of states,
\begin{equation}
N_{A(B)}=\sum_{\gamma} \int_{\Omega_{A(B)}} d^3 {\bbox r} \, \,
{\cal N}^{(A(B))}_{\gamma\gamma}( {\bbox r}),
\end{equation}
 in the region of interest times the fluctuating
potential. Thus the net charge, in region $A$ and $B$, is
\begin{equation}
\hat{Q}_{A}=C_{A} \hat{U}_{A}+C(\hat{U}_A-\hat{U}_B)=e \hat{{\cal
N}}_{A}- e^2N_{A} \hat{U}_{A} , \label{chargeop1}
\end{equation}
\begin{equation}
\hat{Q}_{B}=C_{B} \hat{U}_{B}+C(\hat{U}_B-\hat{U}_A)=e \hat{{\cal
N}}_{B}- e^2N_{B} \hat{U}_{B}. \label{chargeop2}
\end{equation}
Solving Eqs. (\ref{chargeop1},\ref{chargeop2}) gives us, for the
potential operators,
\begin{equation}
    \left(\begin{array}{l}
     \hat{U}_{A} \\
     \hat{U}_{B}
     \end{array} \right)
     =\left(\frac{1}{d} \right)
     \left( \begin{array}{ll}

        [G_B]^{-1}  & C \\

        C & [G_A]^{-1}

        \end{array} \right)
        \left(\begin{array}{l}
     e\hat{{\cal N}}_{A} \\
     e\hat{{\cal N}}_{B}
     \end{array} \right)
        \label{potmat2},
\end{equation}
where $G_{\alpha}=[C_{\alpha} + C + e^2 N_{\alpha}]^{-1}$ and
$d=[G_A]^{-1}[G_B]^{-1}-C^2$. Now that we have an expression for
the potential operators we can calculate the fluctuation spectra
of the internal potentials ${2\pi} S_{U_{\alpha}U_{\beta}}(\omega)
\delta(\omega + \omega^{\prime}) $ $= ({1}/{2})\langle
\hat{U}_{\alpha}(\omega) \hat{U}_{\beta}(\omega^{\prime}) +
\hat{U}_{\beta}(\omega^{\prime}) \hat{U}_{\alpha}(\omega) \rangle,
\nonumber
$
where $\langle..\rangle$ denotes a quantum statistical average
over products of $\hat{a}$ and $\hat{a}^{\dagger}$ operators
\cite{Buttiker2}. From this and using the fact that $S_{Q_{\alpha}
Q_{\beta}}=C_{\alpha} C_{\beta}S_{U_{\alpha} U_{\beta}}$ we obtain
the spectra of the screened charges. For $C=0$ (i.e. no capacitive
coupling between the two regions $\Omega_A$ and $\Omega_B$) we
find
\begin{eqnarray}
S_{Q_{\alpha}Q_{\beta}}(\omega)
=\frac{C_{\mu_{\alpha}}C_{\mu_{\beta}}}{N_{\alpha} N_{\beta}}
\sum_{\gamma \delta} \int dE \, \, F_{\delta \gamma} {\rm Tr}
\left[{\cal N}_{\delta \gamma}^{(\alpha)} ({\cal N}_{\delta
\gamma}^{(\beta)})^{\dagger} \right], \label{chargefluc}
\end{eqnarray}
where
$F_{\delta \gamma}= f_{\delta}(E)[1-f_{\gamma}(E+\hbar \omega)]+$
$f_{\gamma}(E+\hbar \omega)[1-f_{\delta}(E)], $
\begin{equation}
{\cal N}_{\delta \gamma}^{(\eta)}=\int_{\Omega_{\eta}} d^3 {\bbox
r} \, \, {\cal N}^{(\eta)}_{\delta \gamma}({\bbox r}) ,
\label{number}
\end{equation}
and the electrochemical capacitance for a region $\eta$ is
\begin{equation}
C_{\mu_{\eta}}=\frac{e^2N_{\eta}C_{\eta}}{C_{\eta}+e^2 N_{\eta}}.
\label{electro}
\end{equation}
Eq. (\ref{chargefluc}) gives us an expression for the charge
fluctuation spectra, when $\alpha=\beta$, we have the fluctuation
spectra in the gate $\alpha$, when $\alpha \ne \beta$ Eq.
(\ref{chargefluc}) gives us the charge correlations between the
two gates.

At equilibrium, at zero temperature Eq. (\ref{chargefluc}) can be
simplified to $S_ {Q_{\alpha}Q_{\beta}}(\omega)= 2
C_{\mu_{\alpha}} C_{\mu_{\beta}} R_q^{\alpha \beta} \hbar
|\omega|$ where
\begin{eqnarray}
R_q^{\alpha \beta}= \frac{h}{2e^2} \frac{ \sum_{\gamma \delta}
{\rm Tr} \left[ {\cal N}_{\delta \gamma}^{(\alpha)} ({\cal
N}_{\delta \gamma}^{(\beta)})^{\dagger} \right]} {{\rm
Tr}\left[\sum_{\gamma} {\cal N}_{\gamma \gamma}^{(\alpha)} \right]
{\rm Tr} \left[\sum_{\gamma} {\cal N}_{\gamma \gamma}^{(\beta)}
\right]}.
\end{eqnarray}
The other limit we can consider is the zero temperature, low
frequency limit of the charge spectra to leading order in applied
voltage. Evaluating Eq. (\ref{chargefluc}) we find $S_
{Q_{\alpha}Q_{\beta}}(\omega)= 2 C_{\mu_{\alpha}} C_{\mu_{\beta}}
R_V^{\alpha \beta}  |e V|$ where
\begin{eqnarray}
R_V^{\alpha \beta}= \frac{h}{2e^2} \frac{ {\rm Tr} \left[ {\cal
N}_{12}^{(\alpha)} ({\cal N}_{12}^{(\beta)})^{\dagger}
\right]+{\rm Tr} \left[ {\cal N}_{21}^{(\alpha)} ({\cal
N}_{21}^{(\beta)})^{\dagger} \right]} {{\rm Tr}
\left[\sum_{\gamma} {\cal N}_{\gamma \gamma}^{(\alpha)} \right]
 {\rm Tr} \left[\sum_{\gamma} {\cal N}_{\gamma \gamma}^{(\beta)} \right]}.
\end{eqnarray}
From $S_{I_{\alpha} I_{\beta}}=\omega^2 S_{Q_{\alpha} Q_{\beta}}$
we obtain Eqs. (\ref{SII1}) and (\ref{SII2}).

Having constructed the above methodology we now wish to consider a
simple example to see if the current correlations between gates
are positive or negative. The example we choose is a QPC in a high
magnetic field \cite{Buttiker3}, such that the transport is
governed by edge states. We consider four scenarios, as shown in
Fig. 2, keeping gate $A$ fixed and calculating the charge
correlations between gate $A$ and gate $B$ when gate $B$ is in
each of the four separate positions shown. Firstly we consider the
situation where gate $A$ is adjacent to gate $B$ (in position
(i)). These two gates are Coulomb coupled to two separate regions
of the same edge state of interest ($\Omega_A$ and $\Omega_B$).

If we consider only one edge state then the scattering matrix for
this system is
\begin{figure}
\narrowtext \vspace{0.0cm}
\epsfxsize=7cm \centerline{\epsffile{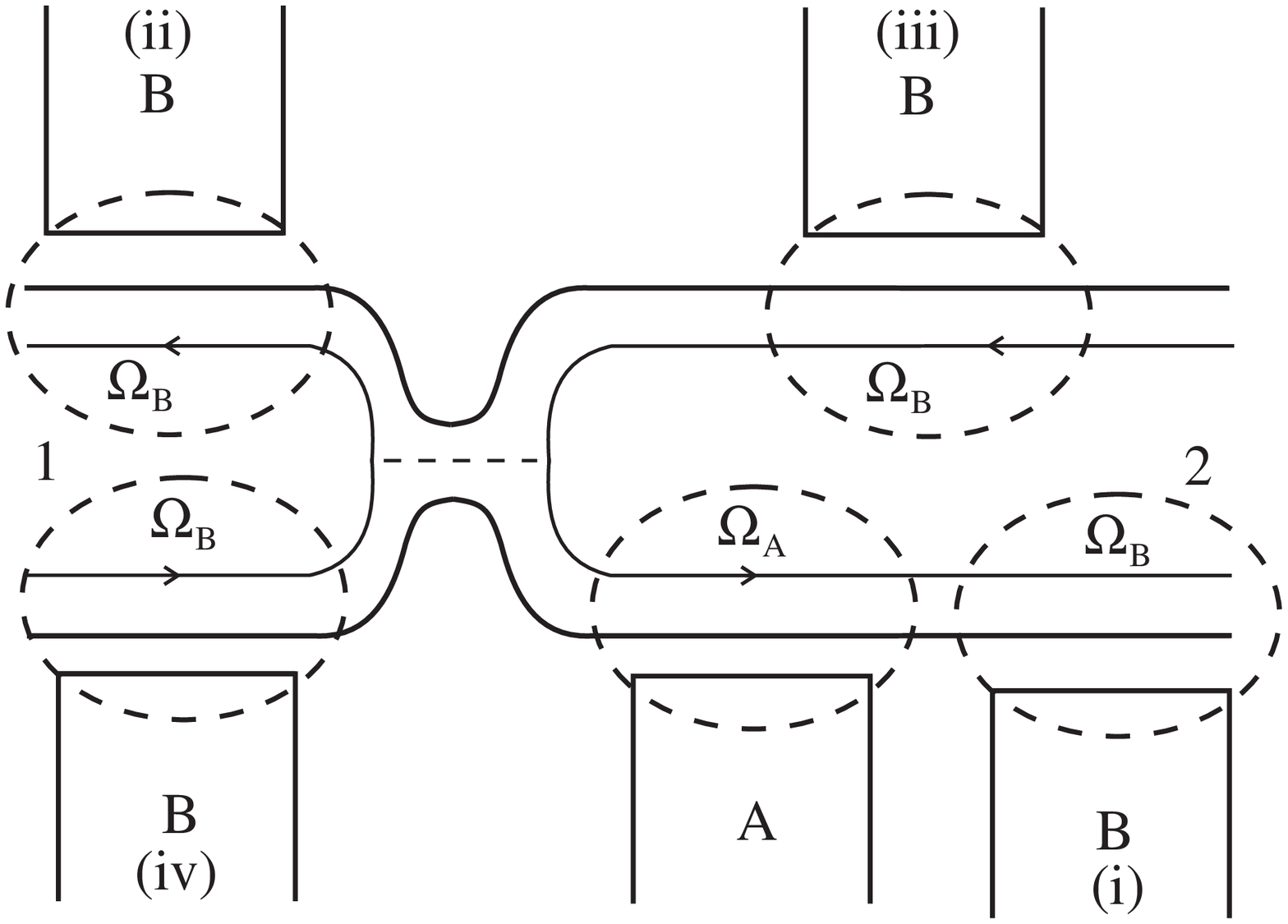}} \vspace{0.3cm}
\caption{ \label{QPC} QPC in a high magnetic field. Gate $A$ is
fixed and the charge correlations between gates $A$ and $B$ for
the four positions ($i$), ($ii$), ($iii$) and ($iv$) of gate $B$
are evaluated. In each case the two gates ($A$ and $B$) are
Coulomb coupled to the edge state which traverses regions
$\Omega_A$ and $\Omega_B$. }
\end{figure}
\begin{equation}
     {\bf S} = \left( \begin{array}{ll}

       r & t  \\

       t^{\prime}e^{(i \theta_A+i \theta_B)} & r^{\prime}e^{(i \theta_A+i\theta_B)}
       \end{array} \right),
        \label{Smat2}
\end{equation}
where the phases, $\theta_A$ and $\theta_B$, have been introduced
explicitly. They arise from the fact that a carrier traversing
region $A$ or $B$ adjacent to either of the gates acquires a
 phase $\theta_A (U_A)$ or $\theta_B
(U_B)$ depending upon the region over which the carrier is
traversing. $U_{A}$ ($U_{B}$) characterizes the potential in
region $A$ ($B$).

To calculate the charge operator we have to evaluate the variation
of the scattering matrix with respect to the potential $U_A$ and
$U_B$ (Eq. (\ref{d0})). Only $s_{12}$ and $s_{22}$ depend upon
these potentials. We find
\begin{equation}
\frac{ds_{\delta \gamma}}{edU_{A(B)}}= \left( \frac{ds_{\delta
\gamma}}{d \theta_{A(B)}} \right) \left( \frac{d
\theta_{A(B)}}{edU_{A(B)}} \right). \label{trans}
\end{equation}
But $(d \theta_{A(B)}/edU_{A(B)})=-2 \pi N_{A(B)}$ where
$N_{A(B)}$ is the density of states of the edge state in region
$\Omega_{A(B)}$. From this we find ${\cal N}_{11}^{(A(B))}  =
N_{A(B)} T$, ${\cal N}_{22}^{(A(B))}  =  N_{A(B)} R$ and
\begin{equation}
     {\cal N}_{12}^{(A(B))}= \left({\cal N}_{21}^{(A(B))} \right)^{\star}=
     N_{A(B)} \left[ (t^{\prime})^{\star} r^{\prime} \right],
\end{equation}
where $T=|t|^2=|t^{\prime}|^2$ and $R=1-T$. Using the fact that
the general scattering matrix for the system can be parameterized
in the following manor
\begin{equation}
{\bf S} = \left( \begin{array}{ll}

       i \sqrt{R} e^{i(\chi+\phi_1)} & \sqrt{T} e^{i(\chi+\phi_2)}  \\

       \sqrt{T} e^{i(\chi-\phi_2)} & i \sqrt{R} e^{i(\chi-\phi_1)}
       \end{array} \right),
        \label{Smat11}
\end{equation}
we have for Eq. (\ref{Smat2})
\begin{eqnarray}
r & =& i \sqrt{R} e^{i(\chi+\phi_1)}, \, \, \,\, r^{\prime}  = i
\sqrt{R}e^{i(\chi-\phi_1-(\theta_A+\theta_B))} \\ t & =&
\sqrt{T}e^{i(\chi+\phi_2)}, \, \,\,\, t^{\prime}  =
\sqrt{T}e^{i(\chi-\phi_2-(\theta_A+\theta_B))}.
\end{eqnarray}
Thus from the above we find
\begin{equation}
R_q^{\alpha \beta}=(h/2e^2) \, \, \, \, \,{\rm and} \, \, \, \, \,
R_V^{\alpha \beta}=(h/e^2)TR. \label{eq:Rv1}
\end{equation}
What we are really interested in is the sign of Eqs. (1,2), we
know for this system that both $R_q^{\alpha \beta}$ and
$R_V^{\alpha \beta}$ are positive, and $C_{\mu_A}$ and $C_{\mu_B}$
are positive, in particular for the structure we have considered
$1/e^2 N_{A(B)} \ll 1/C_{A(B)}$ hence $C_{\mu_{A(B)}} \approx
C_{A(B)}$. From this we deduce that the current correlations
observed between the two gates in this system will be positive and
characterized by Eq. (\ref{eq:Rv1}).

Having considered the example where gate $B$ {\it sees} the same
edge state as gate $A$, it is instructive to consider the case
where gate $B$ is in position $(ii)$. In this case a carrier which
is {\it seen} by gate $A$ will not be {\it seen} by gate $B$ for
this case the scattering matrix is as follows.
\begin{equation}
     {\bf S} = \left( \begin{array}{ll}

       r e^{i \theta_B} & t e^{i \theta_B}  \\

       t^{\prime} e^{i \theta_A} & r^{\prime} e^{i \theta_A}
       \end{array} \right).
        \label{Smat3}
\end{equation}
Following the same methodology as before we find
\begin{eqnarray}
R_q^{AA}&=&R_q^{B B}=(h/2e^2), \\ R_q^{A B}&=&R_q^{B A}=0,
\\
R_V^{AA}&=&R_V^{B B}=-R_V^{A B}=-R_V^{B A}=(h/e^2)TR.
\label{eq:27}
\end{eqnarray}
The equilibrium correlations proportional to $R_q^{A B}$ are zero,
whereas the non-equilibrium correlations given by Eq.
(\ref{eq:27}) are negative.

%
%
%
%

We also consider scenarios ($iii$) and ($iv$) depicted in Fig. 2.
It is possible to calculate $R_q^{AB}$ and $R_V^{AB}$ which
characterize the equilibrium and non-equilibrium current
correlations between the two gates. For clarity we have summarized
our results in table \ref{tb:1} by showing the possible signs of
current correlations between the two gates, for the four possible
gate configurations shown in Fig. 2.

In this work we have shown that through the consideration of
Coulomb coupling macroscopic gates to a mesoscopic conductor it is
possible, for the example we have studied, to gain positive
current correlations between the current fluctuations induced into
the gates. The difference between this result and the result which
would be obtained if we replaced our gates with leads, where the
current correlations would be negative, can easily be understood
by explaining the fundamental difference which arises when only
Coulomb contacts are considered. This difference is that it is
possible for a carrier to be {\it seen} by both contacts when the
gates are only attached via the Coulomb interaction, which is not
the case when we replace the gates with leads where a carrier can
not enter both leads. Consider scenario ($i$) shown in Fig. 2, a
carrier which is {\it seen} by gate $A$ must also be {\it seen} by
gate $B$, replacing the gates with contacts then a carrier
traversing along the edge state of interest can either go into
lead $A$, lead $B$ or continue into lead $2$.

We believe that this is the first prediction of such positive
current correlations in normal mesoscopic conductors. Experimental
work on induced charging in macroscopic gates Coulomb coupled to
mesoscopic conductors has been carried out \cite{chargeing} and we
suggest from this work that induced current fluctuations and
correlations between gates in such systems should be examined in
the future.

This work was supported by the Swiss National Science Foundation
and the TMR network Dynamics of Nanostructures. We thank Ya. M.
Blanter for useful discussions.

\vspace{-0.6cm}
\begin{table}
\begin{center}
\renewcommand{\arraystretch}{1.25}
\begin{tabular}{|c|c|c|c|c|}
\hline
   & ($i$) & ($ii$) & ($iii$) & ($iv$)\\ \hline
  $S^q_{I_A I_B}(\omega)$ & $>0$ & $=0$ & $\geq 0$  & $\geq 0$
\\
\hline $S^V_{I_A I_B}(\omega)$ & $\geq 0$ & $\leq 0$ & $=0$ & $=0$
\\
 \hline
\end{tabular}
\renewcommand{\arraystretch}{1.25}
\vspace{0.3cm} \caption[] {Sign of equilibrium ($S^q_{I_A
I_B}(\omega)$) and non- equilibrium ($S^V_{I_A I_B}(\omega)$)
current correlations between gates $A$ and $B$ for the four
positions of gate $B$ relative to gate $A$. \label{tb:1}}
\end{center}
\end{table}
\end{multicols}
\end{document}